# Defect-based characterization of the fatigue behavior of additively manufactured titanium aluminides

M. Teschke[1,*], J. Moritz[2,3], J. Tenkamp[1], A. Marquardt[2,3], C. Leyens[2,3], F. Walther[1]

*Corresponding author, email: mirko.teschke@tu-dortmund.de, Baroper Straße 303, D-44227 Dortmund, Germany, Phone: +49 231 755 8040, Fax: +49 231 755 8029

[1] TU Dortmund University, Chair of Materials Test Engineering (WPT), Dortmund, Germany

[2] Technische Universität Dresden, Institute of Materials Science (IfWW), Dresden, Germany

[3] Fraunhofer Institute for Material and Beam Technology IWS, Dresden, Germany

**Abstract**

The additively manufactured titanium aluminide alloy TNM-B1 was characterized microstructurally and mechanically in the as-built and hot isostatically pressed (HIP) condition. Tensile and constant amplitude tests were performed at room temperature and 800 °C. Using fractographic SEM images, the fracture-inducing defect was identified. With the HIP, defect number and size could be reduced, increasing fatigue strength by 43% to 500 MPa. Using the model approaches of Murakami and Shiozawa, the fatigue life was correlated with the local stress intensity factor and could be described as function of the stress amplitude as well as the size and location of fracture-inducing defects.

# 1. Introduction

Weight reduction to save energy and reduce $CO_2$ emissions is a major goal in all mobility sectors. Due to their low density (~4 g/cm³), high specific strength, and high temperature resistance [1–3], gamma titanium aluminides (TiAl) have the potential to substitute conventional materials such as nickel-based superalloys, which is of great importance for automotive and aerospace applications [3,4]. Due to their low ductility and fracture toughness at room temperature, machining of TiAl is a great challenge. These alloys are therefore conventionally produced by isothermal forging or centrifugal casting followed by hot isostatic pressing (HIP) and heat treatment, which is expensive, time-consuming, and associated with geometric restrictions [2,5–7]. Thus, alternative manufacturing techniques have been developed and validated. Especially additive manufacturing (AM) techniques are favored since they enable near-net shape manufacturing of highly complex geometries [8,9].

The significance and use of AM technologies for the processing of TiAl alloys has therefore increased continuously in recent years [10]. In this context, powder bed fusion (PBF) processes play an important role. Electron beam powder bed fusion (PBF-EB/M), also known as (selective) electron beam melting (sEBM), has shown the greatest advantages and has been successfully applied in several studies [11–13]. In PBF-EB/M, an electron beam is used to successively preheat the powder bed and melt the cross-sectional area of the part in each layer, followed by lowering the build platform and applying a new layer of powder. The process is explained in detail in [10]. Due to the high process temperature of over 1000 °C, which is above the ductile-brittle transition temperature (DBTT) of TiAl, cracks can be avoided in the manufacturing process [10,14]. The prevailing process conditions prevent oxidation [8,14]. However, depending on the process parameters, process-related defects occur in PBF-EB/M manufactured TiAl-components. If the energy input is too high, material displacement (so-called swelling) and inhomogeneities due to the evaporation of aluminum can occur [15–17]. On the other hand, too low energy input leads to defects such as lack of fusion (LOF) defects (misconnections). Gas trapped in the powder or between powder particles cannot escape due to rapid solidification during melting, resulting in gas porosity [18,19]. Porosity can be reduced by subsequent HIP [20]. Due to the low fracture toughness and damage tolerance, TiAl are particularly sensitive to all forms of defects. These significantly reduce the mechanical strength, hence detailed knowledge of the effect of defects on the quasi-static and cyclic behavior is the key for high performance part manufacturing.

The development of titanium aluminide alloys can be divided into four generations. Due to the low ductility at room temperature of the first generation, the aluminum content of the second generation was increased to 48 at-% in order to significantly increase the room temperature ductility [21,22]. The most well-known representative of this generation is the alloy Ti-48Al-2Cr-2Nb (Ti-48-2-2) with an elongation at fracture of 3% at room temperature, which is already used for low-pressure turbine blades by General Electric [23,24]. The third generation of TiAl alloys was developed to increase the application temperature up to 800 °C. An important representative is the alloy TNM-B1 (Ti-43.5Al-4Nb-1Mo-0.1B), which has been applied in a low-pressure turbine blade into the geared turbofan (GTF) jet engine since 2016 [3,25]. However, this alloy is optimized for casting and forging [6,26] and solidifies via the body-centered cubic β-Ti(Al) phase and not peritectically, as it is usual for many TiAl alloys [14]. The ductility of this alloy is low with a room temperature elongation at fracture after heat treatment of 1.2% for cast components [27]. This alloy has been described in detail by Clemens et al. in [28]. The latest, fourth TiAl generation is currently under development and characterization [29,30]. The focus is on increasing the ductility at room temperature and improving the high-temperature properties as well as the optimization for the PBF-EB/M process.

So far, there have been some studies on the mechanical characterization of AM TiAl, but rarely specifically of the alloy TNM-B1. Most of the investigations have been carried out on conventionally manufactured material, not on AM material. In addition, mechanical characterization is often limited to the determination of quasi-static properties in the tensile test. Extensive quasi-static investigations were carried out on the PBF-EB/M manufactured alloys Ti-48Al-2Nb-2Cr [11,31,32], Ti-47Al-2Nb-2Cr [33], Ti-45Al-8Nb [34] and Ti–47.5Al–5.5Nb–0.5 W [30]. The maximum ultimate tensile strength (UTS) ranged from 353 MPa [11] to 795 MPa [30]. A comprehensive overview is given in [9]. For TNM-B1, on the other hand, there are mainly studies on other manufacturing processes, like spark plasma sintering (SPS) [14] or casting [27]. In SPS, UTS of ~800 MPa (RT) or ~450 MPa (800 °C) are achieved [14], while casting in combination with suitable heat treatments can achieve UTS of 700 to 950 MPa (RT) or 500 to 750 MPa (800 °C) [27]. Teschke et. al [35] investigated the PBF-EB/M manufactured alloy TNM-B1 with a modified chemical composition (specimens built in the z-direction). Due to the stress concentration at internal defects (LOF defects), the UTS was only ~230 MPa (RT) or ~400 MPa (800 °C).

The fatigue behavior of AM TiAl has so far barely been characterized. A general overview of the current state of research in high cycle fatigue (HCF) of TiAl is given in [36]. In [9] a review of fatigue behavior studies specifically for PBF-EB/M manufactured TiAl is given. Most investigations are limited to the PBF-EB/M manufactured alloys Ti-48Al-2Cr-2Nb [20,37–40] as well as the alloy Ti-6Al-4V [41]. For the alloy Ti-48Al-2Cr-2Nb, a fatigue strength (1E7) of 340 MPa (RT) was determined for the stress ratios R = -1 and 0.6. For 700 °C, the fatigue strength was 390 MPa (R = 0.6) for the alloy Ti-48Al-2Cr-2Nb and 416 MPa for the alloy TNM [37,38]. TiAl have a very low defect tolerance due to the low ductility and fracture toughness. The fatigue behavior is therefore very dependent on the size and orientation of the defects as well as the facet size of the lamellae and leads to a very large scatter [9]. Filippi et al. [38] investigated this influence using artificially introduced defects. A modified El-Haddad/Tanka model can describe the relationship between threshold stress range and defect size. Jha et al. [42] found that defects near the surface are significantly more critical to the fatigue behavior of TiAl than internal defects. In different studies, fatigue crack propagation tests were carried out for different manufacturing processes, heat treatments, and stress ratios [20,39,43,44]. The threshold stress intensity range $K_{th}$ varies between 3 and 13 MPa·m$^{1/2}$ depending on the mentioned values.

$$\Delta K_0 = Y \cdot \Delta\sigma \cdot \sqrt{\pi \cdot \sqrt{Area}} = \Delta\sigma \cdot Y \cdot \sqrt{\pi \cdot a_0} \qquad (Eq.\ 1)$$

$\sqrt{Area} = a_0$: Defect size  $\qquad$ *Y*: Geometric factor

Square root of the projected $\qquad$ *Y = 0.65* for *surface* defect

fracture-inducing critical defect $\qquad$ *Y = 0.50* for *internal* defect

Material defects cause local stress concentration and premature failure. In real components, defects and cracks have a complex 3-dimensional geometry, which makes it difficult to measure and describe them. Therefore, Murakami et al. [45,46] modified the equation for the stress intensity factor (SIF) $\Delta K$ (Eq. 1). The formula calculates the cyclic SIF taking into account the stress range $\Delta\sigma$, the defect size $a_0$, which is the square root of the projected critical defect, and the geometry factor Y, which considers the distance from the edge. Defects near the surface, with an edge distance smaller than $a_0$, are defined as surface defects [47,48]. The mentioned models can be used to describe the effect of defects on the fatigue properties [49,50]. So far, no studies are known where these two models have been applied for the evaluation of the fatigue behavior of additively manufactured titanium aluminides with process-related defects.

For crack propagation dominated fatigue behavior of materials, the model of Shiozawa [51] can be used for modified representation and interpretation of fatigue results. Hereby, the Paris-Erdogan law is integrated to describe the crack propagation behavior from the initial crack or defect size $a_0$ to the critical crack length $a_f$ and transformed to the following Eq. 2:

$$\Delta K_0 = \left(\frac{2}{C \cdot (m-2)}\right)^{\frac{1}{m}} \cdot \left(\frac{N_f}{a_0}\right)^{-\frac{1}{m}} = A_1 \cdot \left(\frac{N_f}{a_0}\right)^{A_2}; \; C, m = \text{const.} \quad \text{(Eq. 2)}$$

The Paris coefficient C and exponent m in Eq. 2 can be determined via a power law when representing the cyclic stress intensity factor (SIF) at the fracture-inducing defect $\Delta K_0$ (y-axis) via the quotient of the number of cycles to failure $N_f$ and fracture-inducing defect size $a_0$, the so-called defect-specific number of cycles to failure $N_f/a_0$ (x-axis). With logarithmic scaling, the fatigue results are represented as a straight line analogous to Woehler (S-N) curves.

Due to the lack of data in the characterization of the RT and high temperature quasi-static and fatigue behavior of the additively manufactured alloy TNM-B1, further studies are needed to characterize the alloy and the manufacturing process. In addition, the influence of process-related defects on the mechanical strengths has so far not been taken into account. For these reasons, based on a previous optimization of the process parameters [52] as well as the quasi-static characterization of the initial state [35], the influence of different test temperatures as well as defect sizes (by subsequent HIP) is investigated in this study. The results are evaluated and correlated with the model approaches of Murakami and Shiozawa.

## 2. Material and Methods

## 2.1 Material

In this study, the third-generation TiAl alloy TNM-B1 was investigated. Spherical gas atomized pre-alloyed powder with the nominal chemical composition Ti-46.5Al-4Nb-1Mo-0.1B (in at.-%) and a particle size distribution between 60 and 145 µm (D50 of 83.6 µm) was obtained from GfE Metalle und Materialien [52]. The specimens were manufactured by PBF-EB/M on an Arcam A2X (Arcam AB, Mölndal, Sweden) with the software version EBM Control 3.2. The process parameters are shown in Table 1. These parameters were optimized in terms of porosity and surface quality with the help of a central composite design in a previous study [52]. The specimens were manufactured orthogonally to the building direction (x,y-plane) and then separated by wire EDM followed by turning the specimens into the geometry shown in Figure 1. In the following, the terms and orientations according to DIN EN ISO/ASTM 52900 are specified.

*Table 1: Process parameters for PBF-EB/M process [52].*

| Beam current | Scan speed | Focus offset | Line offset | Layer thickness | Build temp. |
|---|---|---|---|---|---|
| 15 mA | 4000 mm/s | 3 mA | 0.1 mm | 70 µm | > 1000°C |

To reduce the defect size, half of the specimens were hot isostatically pressed (HIP) at 1200 °C/200 MPa for 4 h. Two material conditions were investigated: The as-built (AB) condition and the HIP condition. No further heat treatments were performed.

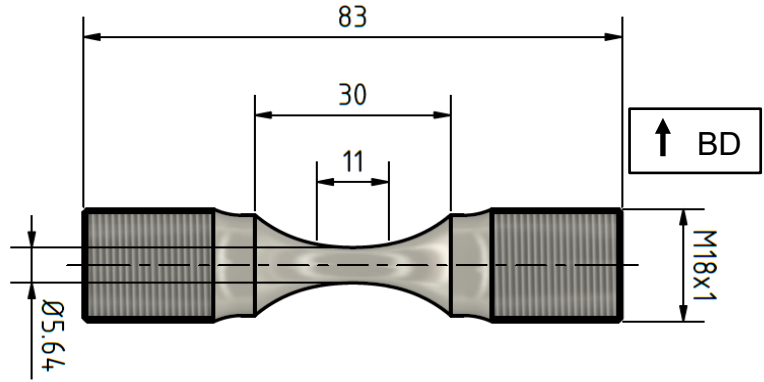

*Figure 1: Specimen geometry for tensile and fatigue tests (BD = build direction).*

## 2.2 Methods

The fatigue and tensile tests were performed on the servohydraulic testing system Instron 8801 (Instron, Norwood, MA, USA), which has a maximum force of $F_{max}$ = 100 kN. Each material condition was tested at room temperature (RT = 20 °C) as well as at 800 °C, which corresponds to the upper range of the application temperature. The two test setups for the two test temperatures are shown in Figure 2. Changes in strain were determined as measurement values for material responses. The strain was recorded with an extensometer Instron 2620-603, with an initial length of $l_0$ = 8 mm at RT and Sandner EXH 10-1A (Sandner Messtechnik, Biebesheim, Germany) with an initial length of $l_0$ = 10 mm for 800 °C. Since the clamping length was too short, the specimens were extended using a self-developed thread adapter made of Inconel. After reaching the test temperature, it was held for approx. 10 minutes and the respective test was then started. Tensile tests were performed with the strain rate of 0.00025 s$^{-1}$. The UTS, the Young's modulus (YM), and the elongation at fracture (EaF) are determined from the stress-strain curves. Stress-controlled constant amplitude tests (CAT) with a sinusoidal load-time function at the stress ratio of R = -1 (fully-reversed loading) and the test frequency of f = 20 Hz were performed. Run-outs were defined as 2E6 cycles, failure represented the complete fracture of the specimen.

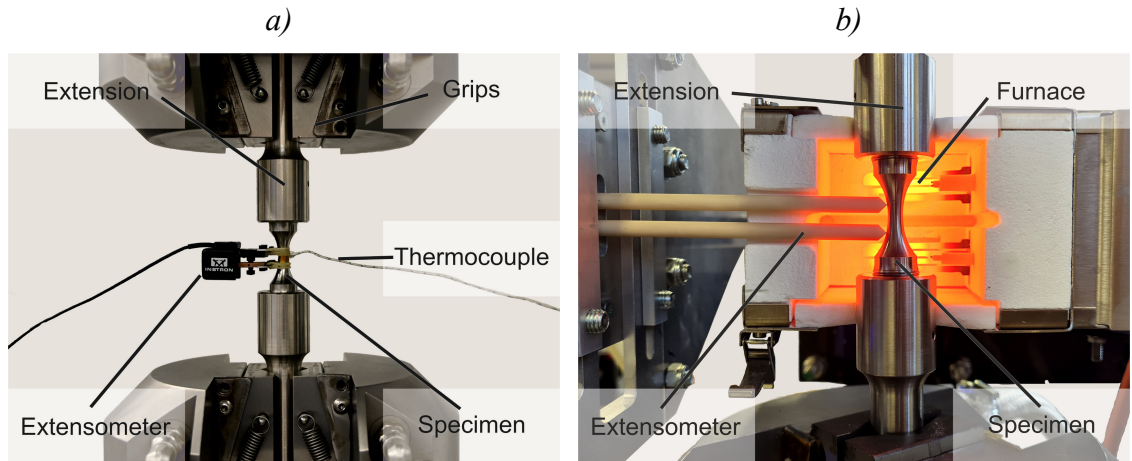

*Figure 2: Test setup for tensile and fatigue tests: a.) Tests at RT; b) Tests at 800 °C. {IN COLOR}*

In microstructural investigations, the initial state was characterized with metallographic cross-sections under SEM Tescan Mira 3 XMU (Tescan, Brno, Czech Republic). In addition, the fracture surfaces of all specimens were examined with the SEM to find the location, size, and shape of each fracture-inducing critical defects. These parameters were determined by measuring the fracture surfaces with the program ImageJ.

The hardness of the two material states was determined by macrohardness measurements on the Wolpert Dia-Testor 2Rc Vickers hardness-testing machine (Instron, Norwood, MA, USA) with a static load of 98.07 N (HV10). For each condition, 10 hardness measurements were performed according to DIN EN ISO 6507-1 and the mean values as well as the standard deviations were determined.

3. **Results and discussion**

3.1 **Microstructural characterization of the initial conditions**

Figure 3 shows the microstructure of the two conditions at the same magnification. The same phases occur in both conditions. The microstructure consists of lamellar $\alpha_2/\gamma$-areas, globular $\gamma$-grains, and the $\beta_O$-phase. However, the HIP resulted in a significant grain coarsening, which led to significant growth of all mentioned phase regions. The hardness of the AB condition is 418 ± 6 HV10, while that of the HIP condition is 385 ± 4 HV10.

This corresponds to a reduction in hardness of -8%. One reason for this is the significant grain coarsening, which is known to result in a reduction of hardness. In addition, the HIP resulted in strong growth of the softer gamma phase fraction, which reduces the integral value of the Vickers hardness.

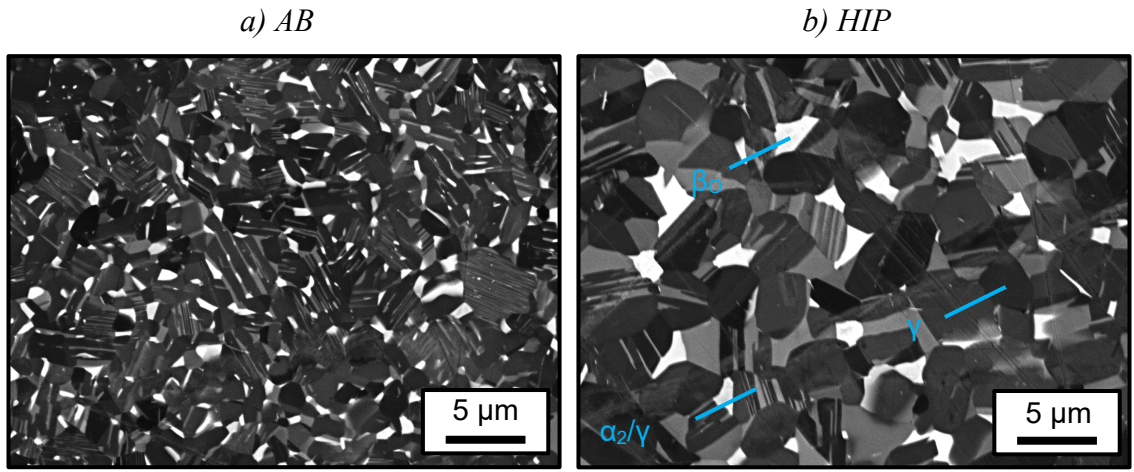

Figure 3: SEM image of PBF-EB/M manufactured TNM-B1 alloy: a) AB and b) HIP condition. {IN COLOR}

## 3.2 Tensile tests

Figure 4 shows a graphical comparison of the UTS, YS, and EaF values from the tensile test for the two test temperatures and material conditions. In Table 2, the individual values are listed. At RT, the UTS significantly increases due to the HIP with +22% to 815 MPa (HIP, RT). At the test temperature of 800 °C, on the other hand, no significant increase in UTS due to the HIP can be seen. However, at $859 \pm 92$ MPa (AB, 800 °C), the UTS is already at a very high level. The reason for this is the increase in ductility due to exceeding the DBTT and consequently the increase in tolerance to internal defects. In investigations by Teschke et al. [35], analogously produced and tested specimens were examined in the AB condition, which, however, were manufactured parallel to the building direction. Due to the unfavorable orientation of the LOF defects, a stress concentration occurred and the UTS was only $226 \pm 17$ MPa (RT) or $400 \pm 33$ MPa (800 °C). By adjusting the building direction, the RT strength could therefore be increased by 195%, the additional HIP improved it by 260% in this investigation. Compared to other TNM-B1 specimens produced by SPS [14] and casting [27], the determined UTS is comparable (RT) or significantly improved (800 °C). Thus, it can be concluded that TiAl components with a high ultimate tensile strength can be produced via PBF-EB/M.

Table 2: Ultimate tensile strength (UTS), Young's modulus (YM), and elongation at fracture (EaF) of PBF-EB/M manufactured TNM-B1 at RT und 800 °C.

| Temperature | Material condition | UTS (MPa) | YM (GPa) | EaF ($10^{-2}$) |
|---|---|---|---|---|
| RT | AB | 666 ± 92 | 185 ± 2 | - |
| RT | HIP | 815 ± 37 | 185 ± 4 | - |
| 800 °C | AB | 859 ± 92 | 153 ± 4 | 0.23 ± 0.12 |
| 800 °C | HIP | 843 ± 20 | 153 ± 9 | 0.67 ± 0.24 |

Since the YM is not significantly affected by defect size, HIP shows no influence on the YM. In contrast, the test temperature, strongly affects the YM, as shown in previous studies, e.g. [35]. Compared to RT (185 GPa), the YM at 800 °C decreases by -17% to 153 GPa. These values are comparable with results from other studies [27,35].

The elongation at fracture could not be determined for the test at RT, since at this temperature the specimens failed before reaching the yield point. This also occurred in previous tests on specimens built parallel to the building direction [35]. The stress concentration at process-related defects led to premature failure due to the low defect tolerance of the material. At 800 °C, however, the elongation at fracture could be measured. Above the DBTT, the material is more ductile and more tolerant to defects at elevated temperatures. The elongation at fracture for the AB condition is 0.23 ± 0.12% and could be increased to 0.67 ± 0.24% by reducing defects with HIP.

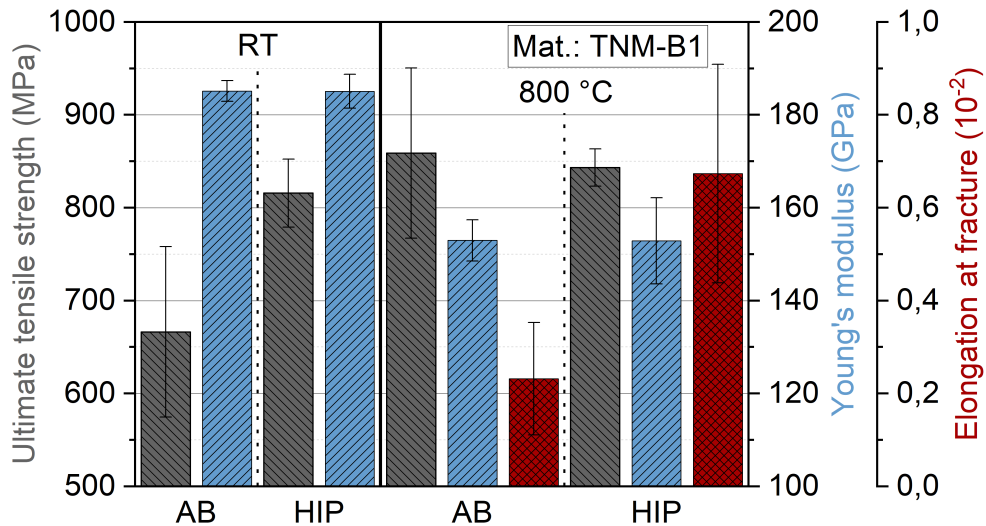

*Figure 4: Ultimate tensile strength (UTS), Young's modulus (YM), and elongation at fracture (EaF) of PBF-EB/M manufactured TNM-B1 at RT und 800 °C. {IN COLOR}*

### 3.3 Fatigue tests

In Figure 5, the stress amplitude and the number of cycles to failure $N_f$, are plotted for both material conditions and test temperatures in a Woehler (S-N) diagram. For both tested conditions, no significant influence of the test temperature on the fatigue behavior could be determined. It can be concluded from this that the material is very well suited to the high operating temperatures in terms of fatigue behavior. In the AB condition, the fatigue strength (2E6) is 350 MPa. In comparison, HIP increased the fatigue strength (2E6) by 43% to 500 MPa. Therefore, it can be assumed that the significant improvement in fatigue strength can be attributed to the HIP by reducing the critical defects. Compared with studies on the PBF-EB/M fabricated alloy Ti-48Al-2Cr-2Nb, where a fatigue strength of 340 MPa (1E7) was achieved, the RT fatigue strength of the AB condition is at a similar level [37,38,53].

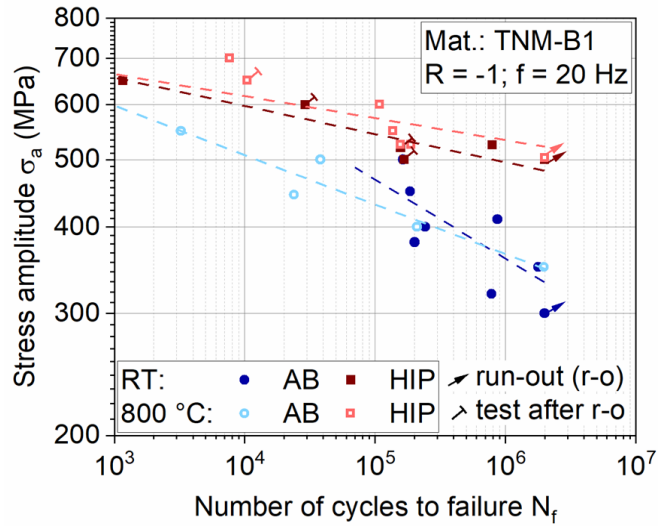

*Figure 5: Woehler (S-N) curves of PBF-EB/M manufactured TNM-B1 at RT und 800 °C. {IN COLOR}*

### 3.4  Fractography

SEM was used to identify the critical defects which lead to failure for each specimen. Different failure mechanisms were identified as the causes of failure. These can be divided into three defect types, which are shown as examples in Figure 6. One mechanism is characterized by LOF between two layers. Since the specimens were extracted orthogonally to the building direction, these defects are cut orthogonally (Figure 6a). A fatigue fracture surface (GBF, granular bright facet) can be detected only for specimens with a SIF threshold value between 6 and 8 MPa·m$^{1/2}$, which could be a threshold value between short and long crack propagation. The threshold value $K_{th}$ has not yet been determined for this alloy. However, based on the results on alloy Ti-48Al-2Cr-2Nb, with $K_{th}$ between 3 and 13 MPa·m$^{1/2}$ [20,39,43,44], this hypothesis should be verified in future investigations taking into account the alloy, the stress ratio, and the test temperature. As another mechanism, failure occurred at pores, often gas porosity (Figure 6d) or unmelted powder particles (Figure 6b). Individual microstructural features were also identified as fracture-inducing (Figure 6c). These were mostly a colony of lamellae with an orientation parallel to the fracture surface which split interlamellarly.

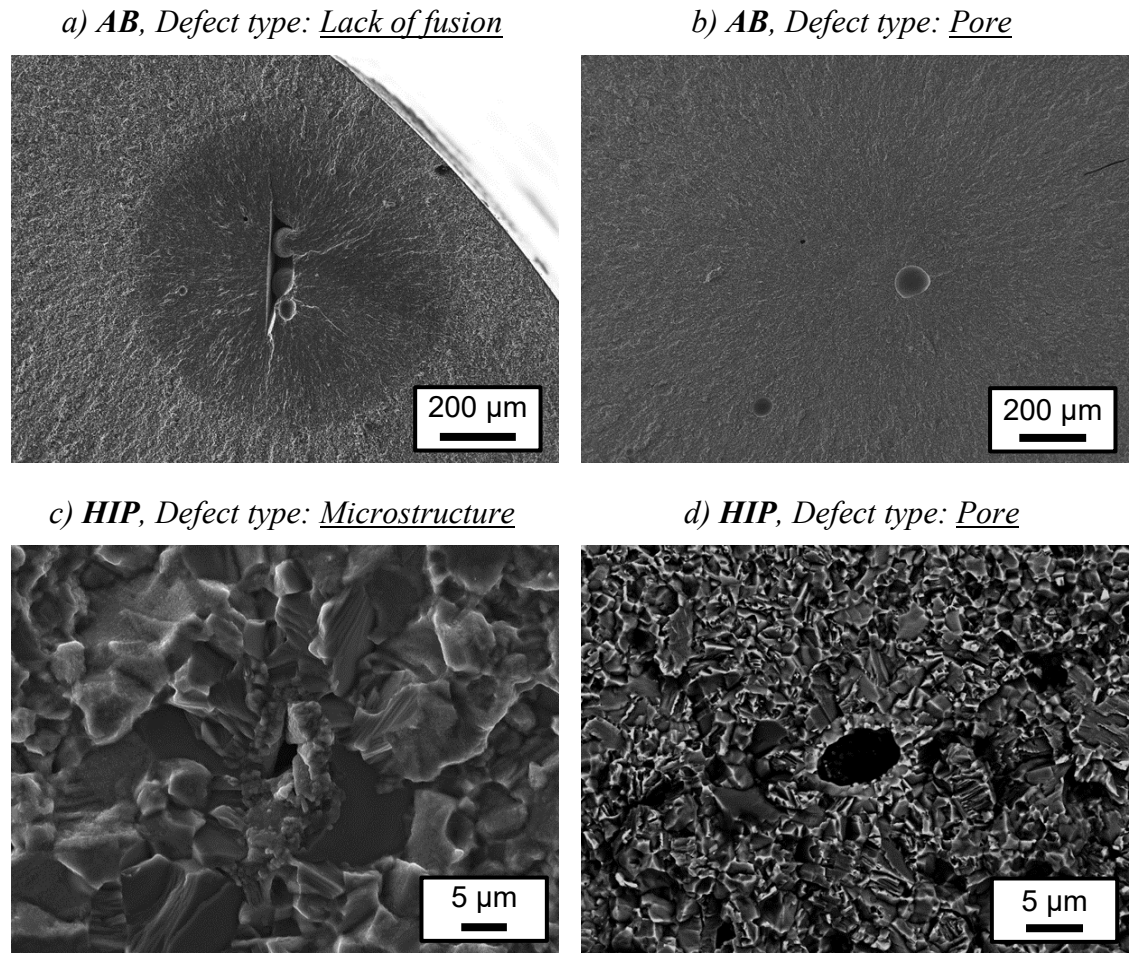

*Figure 6: SEM fractography - Exemplary representation of the three failure mechanisms of PBF-EB/M manufactured TNM-B1 tested in constant amplitude tests at RT and 800 °C. {IN COLOR}*

In Figure 7, the failure mechanism of the identified critical defect which led to failure of each data point is marked in the S-N diagram. This allows to quantitatively assign the dominant failure mechanisms for the two material conditions as well as the two test temperatures. In the AB condition, mainly LOF defects and some gas pores are identified as critical defects. In the HIP condition, gas pores and microstructural defects were the critical defects that led to failure. The fact that LOF no longer played a dominant role in the failure mechanism implies that these defects could successfully be closed by the HIP process. This was possible because the PBF-EB/M process takes place in a nearly vacuum atmosphere. Additionally, the size of the gas pores could be reduced. Different sizes and types of the critical defects cause the different fatigue strengths of the two conditions. Consequently, the S-N diagram is not suitable to describe the relationship between the size of the crack-initiating critical defect, the stress amplitude, and the lifetime. Therefore, another method is required to describe the fatigue behavior, which takes the defect size into account.

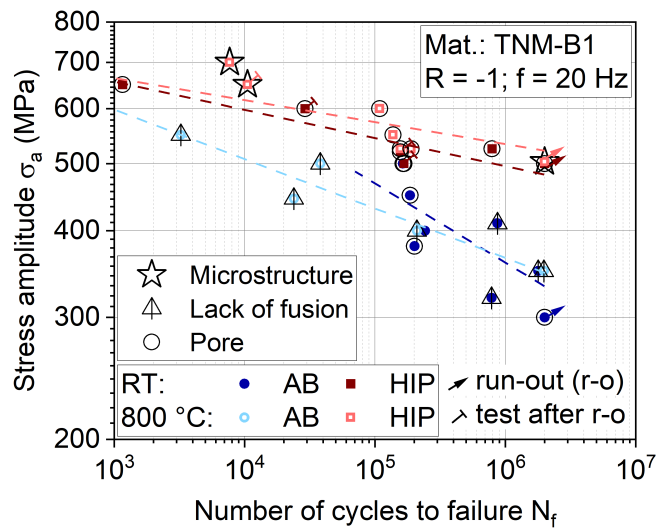

*Figure 7: Woehler (S-N) curves of PBF-EB/M manufactured TNM-B1 at RT und 800 °C. with marked defect types. {IN COLOR}*

### 3.5 Model-based fatigue life prediction

In order to consider the location and size of each defect in the fatigue behavior evaluation, the formula in Eq. 1 is used according to Murakami [45,46]. For each specimen, the critical defect was determined and defect size as well as the stress intensity factor were calculated.

In Figure 8, the defect size of the two material conditions is shown in a boxplot. The different sizes of the critical defects for the respective material conditions are obvious. In the AB condition, the size of the critical defect ranges between 28 and 100 µm (median: 54 µm), whereas after the additional HIP, the defect size could be reduced significantly between 13 and 44 µm (median: 16 µm) due to pore closing and reduction of LOF defects. Moreover, HIP resulted in grain coarsening. Due to the larger lamellae size as well as the elimination of the larger defects, the lamellae in the microstructure also operate as a critical defect in the HIP specimens.

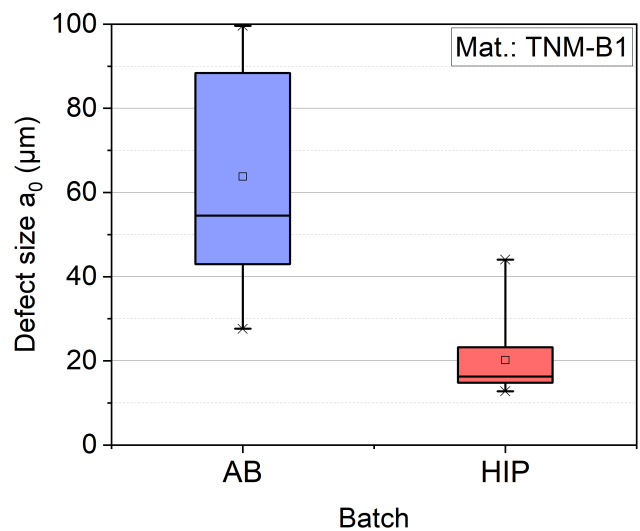

*Figure 8: Defect size of the two PBF-EB/M manufactured material conditions as boxplot determined by SEM fractography. {IN COLOR}*

According to Shiozawa et al. [51,54,55], the calculated cyclic SIF are plotted in the Shiozawa diagram over the normalized defect-specific number of cycles to failure. Despite the different defect sizes both between the two material conditions and within the individual batches, respectively, the relationship between stress amplitude, defect size, and fatigue life can be calculated via a straight line, as can be seen in Figure 9. Using the Shiozawa diagram, it is thus possible to visualize the overall defect-dependent fatigue behavior as a function of the critical defect size for each individual specimen.

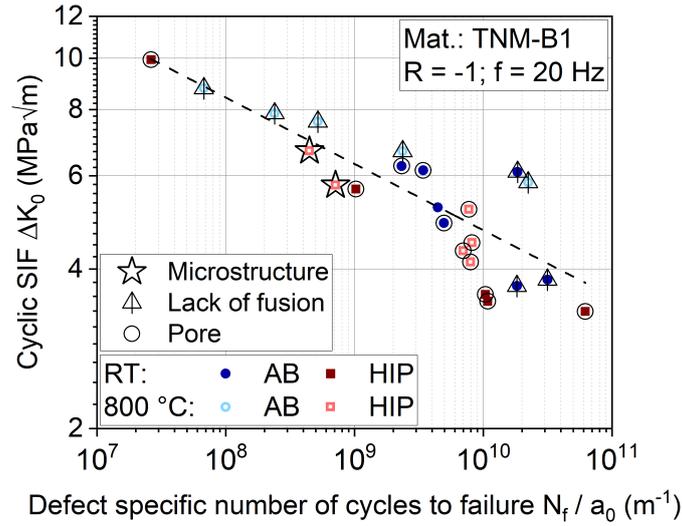

*Figure 9: Shiozawa diagram of PBF-EB/M manufactured TNM-B1 at RT und 800 °C with marked defect types. {IN COLOR}*

The straight line can be described by the formula given in Eq. 2. The Paris coefficient C and exponent m can be determined for the alloy TNM-B1 as C = 1.36E-16 and m = 7.98. Inserting the parameters into the equation gives Eq. 3, which shows the equation for the straight line shown in double logarithmic form for the PBF-EB/M manufactured alloy TNM-B1.

$$\Delta K_0 = 84.78 \cdot \left(\frac{N_f}{a_0}\right)^{-0.125} \quad \text{(Eq. 3)}$$

Despite the good correlation, with an $R^2 = 0.8$, individual outliers can be identified in the HCF range. In future investigations further adjustments to the method will be necessary to take into account other influencing variables that have to be identified, such as the defect shape. In addition, the validity range (e.g. minimum and maximum defect size) of the method must be determined.

If Eq. 1 and 2 are transformed and set equal to each other, Eq. 4 is given, which describes the stress amplitude as a function of the defect size, the geometry factor, and the number of cycles to failure. With Eq. 4, it is possible to calculate synthetic S-N curves for theoretical defect sizes. This makes it possible to describe the effects of the defect size on the S-N curves and thus the fatigue behavior. Exemplary S-N curves for the PBF-EB/M manufactured alloy TNM-B1 are shown in Figure 10. The parameters C and m previously determined for TNM-B1 were used for the calculation, as well as the assumption of internal defects (Y = 0.5).

$$\Delta\sigma(a_0, Y, N_f) = \frac{\left(a_0 \cdot \frac{2}{C \cdot (m-2)}\right)^{\frac{1}{m}}}{Y \cdot \sqrt{\pi \cdot a_0}} \cdot N_f^{-\frac{1}{m}} \qquad \text{(Eq. 4)}$$

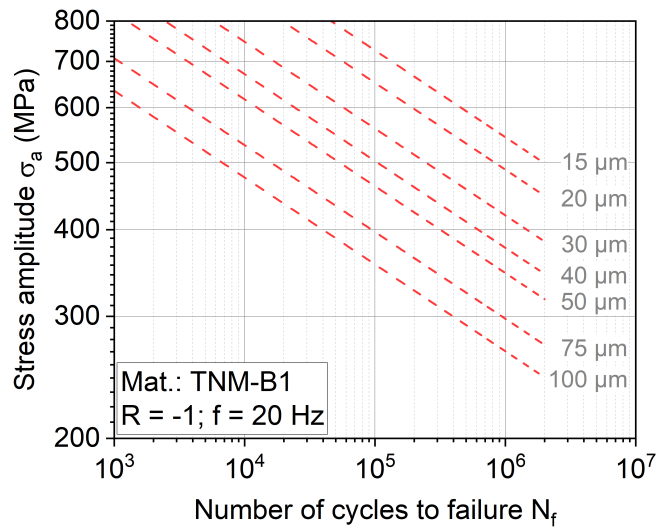

Figure 10: Synthetic S-N curves for theoretical defect sizes based on (Eq. 4) with C = 1.36E-16, m = 7.98 and Y = 0.5 for PBF-EB/M manufactured TNM-B1. {IN COLOR}

With the help of synthetic S-N curves, it can be decided what defect size is acceptable to meet the requirements of a specific application. The constants C and m need to be determined only once for a material. Subsequently, the S-N curves can be generated without a constant amplitude test if the defect size and position of the critical defects are known. This could be possible with computer tomography or within a process monitoring. Possible applications for this are the rapid component design and quality control, where a decision can be made whether or not the component meets the requirements based on occurring defects.

## 4. Conclusions and Outlook

In this study, the titanium aluminide alloy TNM-B1, manufactured by electron beam powder bed fusion orthogonally to the building direction, was characterized in the as-built condition and after hot isostatic pressing. The two material conditions were characterized by means of metallographic methods, tensile tests, and fatigue tests at room temperature and 800 °C.

The elevated test temperature resulted in a significant increase in ultimate tensile strength as well as ductility, with the highest values obtained for the hot isostatic pressed condition. In contrast, the test temperature has no significant influence on fatigue strength. The additional hot isostatic pressing increased the fatigue strength (2E6) by 43%. The fatigue strength is significantly affected by the size of the defects. In the as-built condition, lack of fusion defects and pores are the critical defects that lead to failure. With hot isostatic pressing, the defect size could be significantly reduced, pores and microstructural defects are critical defects here.

To describe the lifetime as a function of the defect size, the models of Murakami and Shiozawa were used. This correlation was valid for both material conditions and test temperatures. It is possible to calculate synthetic S-N curves for theoretical defect sizes and thus evaluate the influence of different defects.

In future investigations, the influence of different microstructures (heat treatments) on the fatigue behavior and the defect model should be investigated. In addition, the defect model should be extended by further influencing variables to be identified (e.g. defect shape) and the application limits should be determined.

**Declaration of Competing Interest**

The authors declare that they have no known competing financial interests or personal relationships that could have appeared to influence the work reported in this paper.

**Acknowledgment**

The authors thank the German Research Foundation (Deutsche Forschungsgemeinschaft, DFG) for its financial support within the research project "Microstructure and defect controlled additive manufacturing of gamma titanium aluminides for function-based control of local materials properties" (project number: 404665753).